\documentclass{mnras}
\usepackage{graphicx}
\usepackage{txfonts}
\let\bibhang\relax
\usepackage{natbib}
\bibpunct{(}{)}{;}{a}{}{,}

\title[Millimetre emission of six debris disks]{New constraints on the millimetre emission of six debris disks}
\author[J.~P. Marshall et al.]{Jonathan P. Marshall$^{1,2,3,}$\thanks{E-mail: jonty.marshall@unsw.edu.au}, S.~T. Maddison$^{4}$, E. Thilliez$^{4}$, B.~C. Matthews$^{5,6}$, 
\newauthor D.~J. Wilner$^{7}$, J.~S. Greaves$^{8}$, W.~S. Holland$^{9,10}$ \\
$^{1}$School of Physics, UNSW Australia, High Street, Kensington, Sydney, NSW 2052, Australia\\
$^{2}$Australian Centre for Astrobiology, UNSW Australia, High Street, Kensington, Sydney, NSW 2052, Australia\\
$^{3}$Computational Engineering and Science Research Centre, University of Southern Queensland, Toowoomba, QLD 4350, Australia\\
$^{4}$Centre for Astrophysics and Supercomputing, Swinburne University of Technology, Hawthorn, VIC 3122, Australia\\
$^{5}$National Research Council of Canada, 5071 West Saanich Rd, Victoria, BC V9E 2E7, Canada\\
$^{6}$Department of Physics and Astronomy, University of Victoria, Victoria, BC V8W 2Y2, Canada\\
$^{7}$Harvard-Smithsonian Center for Astrophysics, 60 Garden Street, Cambridge, MA 02138, USA\\
$^{8}$School of Physics and Astronomy, Cardiff University, Queen's Buildings, The Parade, Cardiff CF24 3AA, UK\\
$^{9}$UK Astronomy Technology Center, Royal Observatory, Blackford Hill, Edinburgh EH9 3HJ, UK\\
$^{10}$Institute for Astronomy, University of Edinburgh, Royal Observatory, Blackford Hill, Edinburgh EH9 3HJ, UK}

\begin{document}

\date{Accepted ---. Received ---; in original form ---}

\pagerange{000--000} \pubyear{---}

\maketitle

\label{firstpage}

\begin{abstract}
The presence of dusty debris around main sequence stars denotes the existence of planetary systems. Such debris disks are often identified by the presence of excess continuum emission at infrared and (sub-)millimetre wavelengths, with measurements at longer wavelengths tracing larger and cooler dust grains. The exponent of the slope of the disk emission at sub-millimetre wavelengths, `$q$', defines the size distribution of dust grains in the disk. This size distribution is a function of the rigid strength of the dust producing parent planetesimals. As part of the survey `PLAnetesimals around TYpical Pre-main seqUence Stars' (PLATYPUS) we observed six debris disks at 9-mm using the Australian Telescope Compact Array. We obtain marginal ($\sim$~3-$\sigma$) detections of three targets: HD~105, HD~61005, and HD~131835. Upper limits for the three remaining disks, HD~20807, HD~109573, and HD~109085, provide further constraint of the (sub-)millimetre slope of their spectral enery distributions. The values of $q$ (or their limits) derived from our observations are all {smaller} than the oft-assumed steady state collisional cascade model ($q = 3.5$), but lie well within the theoretically expected range for debris disks $q \sim 3$ to 4. The measured $q$ values for our targets are all $<$ 3.3, consistent with both collisional modelling results and theoretical predictions for {parent planetesimal bodies being `rubble piles' held together loosely by their self-gravity}. 
\end{abstract}


\section{Introduction}

The debris disks observed around main sequence stars are the remnants of primordial proto-planetary disks \citep{2011WilCie,2015Wyatt}, their constituent dust produced through the collisional attrition of {asteroidal and cometary} bodies, or {deposition by sublimation from comets} \citep{2008Wyatt,2014Matthews}. The size distribution of dust grains in a disk is observationally estimated by measuring the slope of the sub-millimetre emission from the disk {\citep{1990Beckwith,2012Gaspar}}. This slope is a function of the dust emissivity, stellar luminosity \citep{1979BLS}, and the collisional state of the disk \citep{2010Krivov}.

Collisional models of disks produce different predictions for the grain size distribution depending on the dynamical state or physical condition of the colliding planetesimals. In the steady state scenario, a collisional cascade results from inelastic collisions and fragmentation that results in the dust grains following a power law size distribution, i.e. $\delta n \propto a^{-q} \delta a$, with a $q$ of 3.5 \citep{1969Dohnanyi}. That standard model assumes a single constant tensile strength and velocity dispersion of the colliding bodies independent of their size. Such a scenario is unlikely to be true in nature. 

In the Solar system, the size distribution of the largest bodies is a two component power law with a break at a radius of $\sim$~100~km \citep{2004Bernstein,2006Bernstein,2009FraKav,2008FueHol,2009Fuentes}. Above this radius the bodies have not been subject to disruptive collisions and the size distribution exponent, $q~=~5$, is steeper than below it where $q~=~3$. For bodies below the break radius the size distribution exponent is representative of collisional population whose binding energy is dominated by gravity \citep{2005PanSar}. At smaller radii, around 0.1 to 1~km, material strength dominates over gravitational binding energy and a different $q$ applies to the size distribution of those bodies \citep{1999BenAsp}. However, direct observation of Edgeworth-Kuiper belt bodies $<$~1~km in size is difficult due to the faintness of these bodies with typical albedoes $\sim$ 6 per cent. For the smallest grains, a size distribution exponent of $q~=~3$ is consistent with the non-detection of thermal emission from the disk \citep{2012Vitense}, although smaller $q$ values cannot be ruled out by the available data.

The assumed physical condition of dust parent bodies has a large impact on the derived size distribution of dust grains, varying from 3 to 4 for most models. As above, treating the dust-producing planetesimal bodies as `rubble piles' held together by gravity with little internal strength produces $q$ values around 3.0 to 3.3 \citep{2005PanSar}. The collisional modelling code `Analysis of Collisional Evolution' \citep{2013Krivov}, with size dependent grain strength, orientation, and mutual gravity between colliding bodies, produces $q$ values $\sim$ 3.3 to 3.4 \citep{2012Lohne,2014Schuppler,2015Schuppler}. {Including (one or more) refinements to the physics of the collisional modelling such as size-dependent tensile strength laws, mutual gravitation, and size-dependent velocity distributions resulting from viscous stirring and collisional damping \citep{2004KenBro,2005PanSar,2008Lohne,2012Gaspar,2012PanSch} tends to produce steeper grain size distributions (higher $q$) of up to 3.8 to 4 when the tensile strength or velocity dispersion of the colliding bodies are steep functions of their size.}

The theoretical expectation that $q$ will lie in the range 3 to 4 is {borne} out for the majority of disks. Observations at millimetre wavelengths using the Green Bank Telescope \citep{2012Greaves}, the Australian Telescope Compact Array \citep{2012Ricci,2015Ricci}, and the Jansky Very Large Array \citep{2016Macgregor} demonstrated that most systems fell within the range of $3.0 < q < 4.0$. At far-infrared wavelengths, \cite{2014Pawellek} modelling of 34 resolved debris disks gave an error-weighted mean value of $q~=~$3.92~$\pm$~0.07, whilst at millimetre wavelengths \cite{2016Macgregor} obtain an error-weighted value for $q~=~$3.36~$\pm$~0.02 from a sample of 15 disks. However, a number of pathological cases have been identified. For these disks the values of $q$ determined from modelling of their spectral energy distributions (SEDs) lie well beyond the range of theoretically expected values. Such `steep SED' disks identified at far-infrared and sub-millimetre wavelengths, e.g. \cite{2012Ertel}, \cite{2016Montesinos}, and {\cite{2016Marshall}}, have $q$ values between 5 and 10, much higher than theoretical predictions.

The underlying nature of these disks with steep slopes is an open question. To better address this, determination of the slope values exhibited by disks across a broad range of spectral types and ages is required. Here we report results from PLAnetesimals around TYpical and Pre-main seqUence Stars (PLATYPUS), an ongoing survey of debris disks at 9 mm with the Australian Telescope Compact Array (ATCA). {The targets in this work are six debris disk host stars spanning a broad range of stellar luminosities and ages}.

In Section \ref{sec:Obs} we present the new observations made for the PLATYPUS survey, along with a summary of our data reduction and image analysis procedures. In Section \ref{sec:Res} we present the results of our analysis, and combine these six targets with results from previously published phases of the PLATYPUS survey to examine the whole sample. In Section \ref{sec:Dis} we discuss the survey results, both for individual targets and in aggregate. Finally, we present our conclusions and recommendations for future work in Section \ref{sec:Con}.

\section{Observations and analysis}\label{sec:Obs}

\subsection{Observations}

\begin{table}
 \caption{Stellar physical parameters. \label{tab:TgtProps}}
 \centering 
 \begin{tabular}{lcccccr}
  \hline
  Target &     R.A.        & Dec.        &   $d$    & Sp.  &  Age  & Ref. \\
         &    (h m s)      &    (d m s)  &   (pc)   & Type & (Myr) &      \\  
  \hline\hline
  HD~105    & 00 05 52.54 & -41 45 11.0 &  39.4 & G0~V &   27 & 1,2 \\
  HD~20807  & 03 18 12.82 & -62 30 22.9 &  12.0 & G0~V & 2000 & 3 \\
  HD~61005  & 07 35 47.46 & -32 12 14.0 &  35.3 & G8~V &   90 & 4 \\
  HD~109085 & 12 32 04.22 & -16 11 45.6 &  54.7 & F2~V & 1500 & 5 \\
  HD~109573 & 12 36 01.03 & -39 52 10.3 &  72.8 & A0~V &    8 & 6 \\ 
  HD~131835 & 14 56 54.47 & -35 41 43.7 & 122.7 & A2~V &   17 & 7 \\ 
  \hline
 \end{tabular}

\medskip
\raggedright
{References:} 1. \cite{2000Torres}; 2. \cite{2000ZucWeb}; 3. \cite{2013Eiroa} ; 4. \cite{2011Desidera}; 5. \cite{2003Malik} ; 6. \cite{1995Stauffer} ; 7. \cite{2015Moor}.
\end{table}

A search for continuum emission at 9 mm from six debris disks was undertaken by ATCA using the Compact Array Broadband Backend digital filter bank \citep{2011Wilson}. These observations comprise the second part of the PLATYPUS survey (program ID C2694, P.I. Maddison), the previous results having been presented in \cite{2012Ricci,2015Ricci}. The targets selected here by-and-large are older and host fainter disks than those targets previously examined in comparable works \citep{2015Ricci,2016Macgregor}. A summary of relevant stellar parameters is given in Table \ref{tab:TgtProps}. Positions are the stellar optical position in the ICRS at J=2000 epoch. Distances were taken from the re-reduction of the Hipparcos catalogue \citep{2007vanL}. Ages were taken from the literature.

Predicted 9-mm flux densities for the targets ranged from 50 to 150~$\mu$Jy, and were based on modified blackbody models fitted to available far-infrared and sub-millimetre photometry. {The integration times were between 290 and 360 mins}, equivalent to an rms of 15~$\mu$Jy/beam, chosen to ensure detection of the disks' emission based on extrapolations from available measurements. Apart from HD~61005, which has been observed at 9 mm \citep{2016Macgregor}, none of the targets have flux density measurements beyond 1.3 mm. This is especially important for HD~105 and HD~20807, whose sub-millimetre SEDs are particularly poorly constrained. As an additional complication, HD~109085 hosts a large, extended disk and the total disk emission might therefore be resolved out with the array configuration used in these observations. The remaining disks were expected to be point-like at the resolution of our ATCA observations. 

\begin{table}
 \caption{Observation log. \label{tab:ObsLog}}
 \centering 
 \begin{tabular}{llllll}
  \hline
  Date & Target & & Calibrators & & $t_{\rm obs}$ \\
  (2015)     &        & Bandpass & Flux & Phase &  (mins)\\
  \hline\hline
  Apr 4 & HD~109573 & 0537-441  & 1934-638 &  1144-379 & 350 \\
  Apr 5 & HD~109085 & 0537-441  & 1934-638 &  1213-172 & 330 \\
  Apr 6 & HD~131835 & 1424-418  & 1934-638 &  1451-375 & 300 \\
  Jul 13 & HD~61005 & 0537-441  & 1934-638 &  0745-330 & 310 \\
  Jul 14 & HD~105   & 1613-586  & 1934-638 &  2326-477 & 290 \\
  Jul 15 & HD~20807 & 1921-293  & 1934-638 &  0308-611 & 390 \\
  \hline
 \end{tabular}
\end{table}

Observations were taken in two three-day spans covering 2015 April 4 to 6 (HD~109573, HD~109085, and HD~131835) and 2015 July 13 to 15 (HD~61005, HD~105, and HD~20807). All targets were observed using the compact H214 configuration, with baselines between 82 and 247~m, giving a synthetic beam FWHM of $\sim$ 5\arcsec~$\times$~4\arcsec. This angular resolution is sufficient to resolve the nearest/largest disks in the sample, given sufficient signal-to-noise to interpret the emission. The ATCA comprises six 22-m antennae, one of which is located at a fixed position 6 km from the others. In our analysis we have opted not to include data from the fixed antenna due to the increased phase noise on the much longer baselines. A summary of the observations is presented in Table \ref{tab:ObsLog}.

\subsection{ATCA data reduction}

The ATCA observations were reduced in the {\sc Miriad} software environment (version 1.5), following the recommended protocols for millimetre data as given in Chapter 4 of the ATCA Users Guide\footnote{https://www.narrabri.atnf.csiro.au/observing/users\_guide/html/atug.html}. Flagging of radio frequency interference was undertaken using the package {\sc pgflag} in automated mode. The degree of RFI varied strongly with the baseline under consideration and the date of observation, but the fraction of discarded data was around 15 to 30 per cent. We used the standard packages {\sc invert}, {\sc mfclean}, and {\sc restor} to create images from each dataset. Initially, the two sidebands taken of each target (2 GHz bandwidth centred on 33 and 35 GHz) were reduced separately to examine each individually for reduction artefacts (e.g. from phase errors) or elevated noise. Finding none, we combined the two sidebands into a single map with a reference frequency of 34 GHz (9 mm) to reduce the overall noise of the final image. Source flux densities and uncertainties were measured using the package {\sc imfit}, which fits a reconstructed PSF to a given position in the image and derives the peak flux and residual noise. 

\section{Results}\label{sec:Res}

\begin{figure*}
\centering
\includegraphics[width=\textwidth]{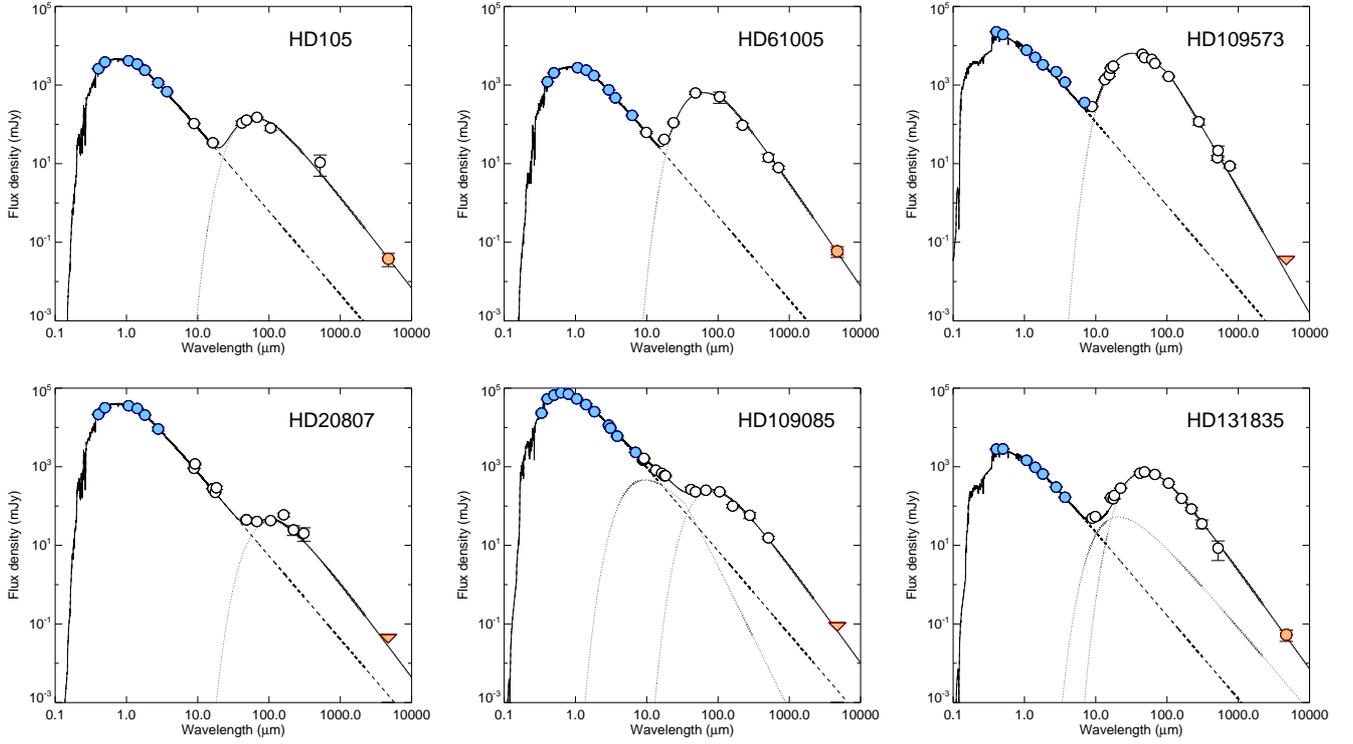}
\caption{SEDs of the six debris disks examined in this work. White dots are ancilliary optical and infrared photometry, as listed in Table \ref{tab:Phot}. The blue dots denote photometry used to scale the stellar photosphere for each system. The red dot/triangle denotes the ATCA 9 mm flux/3-$\sigma$ upper limit. The dashed grey line denotes the stellar photosphere, whilst the dotted grey line denotes the disk contribution(s). The black solid line is the total star+disk model.}\label{fig:SED}
\end{figure*}

\begin{figure*}
\centering
\includegraphics[width=\textwidth,trim={3cm 0 3cm 0},clip]{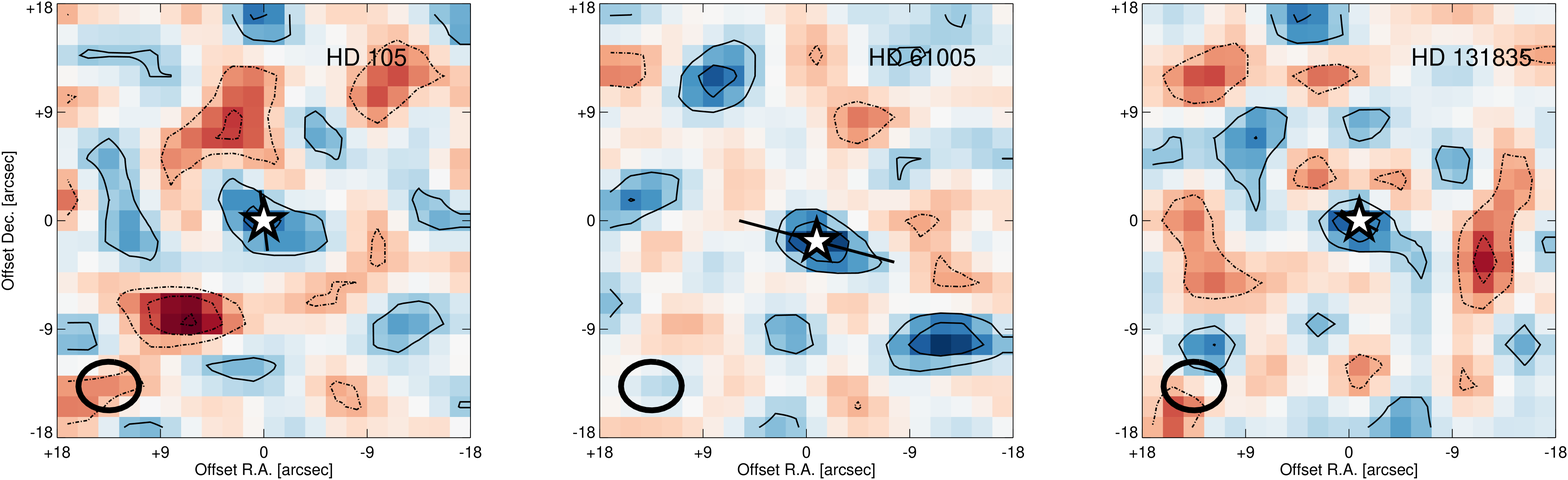}
\caption{Images of the three debris disks detected in our 9-mm ATCA observations, \textit{left to right} HD~105, HD61005, HD~131835. The stellar position is denoted by the `$\star$' symbol. {The thick solid lines denotes the disk extents and position angles, taken from \textit{Herschel} (HD~105) or scattered light (HD~61005, HD~131835) imaging observations \citep{2012Donaldson,2007Hines,2015Hung}}. {The extension of HD~61005's disk along the expected position angle is not significant, given the low s/n of the detection.} Contours are in 1-$\sigma$ increments from $\pm$~2-$\sigma$ and dotted lines denote negative flux. Colour scale is linear from -3-$\sigma$ to +3-$\sigma$. The ATCA beam ($5\arcsec\times3\arcsec$) is illustrated by the ellipse in the bottom left corner. Orientation is north up, east left.}\label{fig:img}
\end{figure*}

\begin{table}
\caption{ATCA photometric measurements. \label{tab:Phot}}
\centering
\begin{tabular}{lccc}
\hline
Target & Beam size & Beam P.A. & $F_{\rm 9 mm}$$^{a}$ \\
       & (\arcsec) &  (\degr)  &   ($\mu$Jy/beam)    \\
\hline\hline
HD~105    &  5.0~$\times$~3.9  & 79.4 &  42~$\pm$~14 \\
HD~20807  &  5.2~$\times$~4.4  & 76.5 &       $<$~54 \\
HD~61005  &  5.1~$\times$~3.9  & 81.3 &  59~$\pm$~13 \\
HD~109085 &  5.4~$\times$~4.0  & 79.3 &       $<$~36 \\
HD~109573 &  5.1~$\times$~3.9  & 79.2 &       $<$~63 \\
HD~131835 &  5.0~$\times$~3.8  & 80.9 &  53~$\pm$~17 \\
\hline
\end{tabular}

\medskip
\raggedright

{Notes:} $^{a}$Upper limits are 3-$\sigma$.
\end{table}

We combined the newly acquired ATCA 9-mm photometry with observations from the literature spanning optical to millimetre wavelengths to produce SEDs for analysis. The optical data come from the Hipparcos catalogue \citep{1997Perryman}; near-infrared photometry are taken from the 2MASS \citep{2003Cutri} and AllWISE \citep{2010Wright} catalogues; mid- and far-infrared photometry were taken by \textit{Spitzer} and \textit{Herschel} \citep{2012Donaldson,2013Eiroa,2014Duchene,2015Moor,2016Morales}; the (sub-)millimetre data were from APEX, JCMT, and SMA \citep{2010Nilsson,2013Ricarte,2016Macgregor}. A summary of the ATCA photometry used in the disk modelling is presented in Table \ref{tab:Phot}. 

The target spectra are presented in Fig. \ref{fig:SED}. The stellar component of the model was estimated by scaling a Castelli-Kurucz photospheric model \citep{2004CK} of appropriate spectral type to optical and near-infrared photometry up to 10~$\mu$m, typically the $BVJHK_s$ and W3 measurements. The disk contribution to the total flux was treated as a single temperature modified blackbody with a break wavelength $\lambda_{\rm 0}$ and exponent $\beta$ such that up to $\lambda_{\rm 0}$ the emission was blackbody-like, but beyond $\lambda_{\rm 0}$ the blackbody emission was {multiplied} by a factor $(\lambda_{\rm 0}/\lambda)^{\beta}$. The disk temperature, $\lambda_{\rm 0}$, and $\beta$ were determined from a least-squares fit to the {star-subtracted} photometry beyond 50$\mu$m weighted by the uncertainties. For HD~109085 and HD~131835, an additional warm blackbody component was used to fit the mid-infrared emission prior to fitting the cold component, adopting the properties taken from \cite{2014Duchene} and \cite{2015Moor}, respectively. 

Having fitted the target SEDs we then proceeded to derive the exponent of the dust grain size distribution $q$ following \cite{2016Macgregor}. To summarise briefly, we calculated the slope of a Planck function, $\alpha_{\rm Pl}$, from our observations at 9 mm and the next longest wavelength, typically 850~$\mu$m or 1.3~mm: 

\begin{equation} 
  \label{Eqn:aPl}
  \alpha_{\rm Pl} = 2 + \frac{ \log \left( \frac{ 2 k T_{\rm d} - h\nu_{1} }{ 2 k T_{\rm d} - h\nu_{2} } \right) } {\log(\nu_{1}/\nu_{2})}
\end{equation}

\noindent $k$ is the Boltzmann constant, $h$ is the Planck constant, $T_{d}$ is the disk temperature, and $\nu_{1,2}$ are the frequencies of our two (sub-)millimetre observations. We also calculated the slope of the millimetre emission, $\alpha_{\rm mm}$ in similar fashion:

\begin{equation} 
  \label{Eqn:amm}
  \alpha_{\rm mm} = \frac{ \log(F_{\nu_{1}}/F_{\nu_{2}}) }{ \log(\nu_{1}/\nu_{2}) }
\end{equation}

\noindent $F_{1,2}$ are the disk fluxes at the two frequencies of interest, $\nu_{1,2}$. {Eqn. \ref{Eqn:amm} assumes that both measurements used to derive the slope are in the Rayleigh-Jeans limit}, which is valid for our debris disks. The value of $q$, the exponent of the grain size distribution, can be derived from these two quantities through the relation: $q = (\alpha_{\rm mm} - \alpha_{\rm Pl})/\beta_{s} + 3$. The relation holds for values of $q$ in the expected range of 3 to 4 that is typical for debris disks, and so should be valid here. {In this relation the $\beta_{s}$ parameter is taken from \cite{2006Draine} who derived the relation between $\beta$ and $q$ for grains in protoplanetary disks. The value of $\beta_{s}$ is dependent on the assumed composition with a range of 1.3, from material produced by the pyrolisis of cellulose at 800~$\degr$C \citep["cel800"][]{1994Jager}, to 2.0, derived for simple models of insulators and conductors \citep{2004Draine}. Whilst values outside this range have been identified in specific cases \citep[e.g.][]{1978Bosch,1996Agladze} their applicability to isolated small particles are not certain; small spheres are expected to have $\beta_{s} \geq 1.5$, but we here adopt a range $\beta_{s}$ of 1.3 to 2.0 as a conservative approach \citep[see Sect. 2][]{2006Draine}. For astronomical silicate, an oft-assumed debris dust material, the appropriate $\beta_{s}$ value is 1.8 and we use this value to calculate the $q$ values presented here. The range of $q$ for a disk resulting from selection of different $\beta_{s}$ values may be treated as a systematic uncertainty in the value of $q$, and are given separately in the table.} The results of the disk fitting process and dervied $q$ values are presented in Table \ref{tab:qvals}.

{Using this relation we can also calculate the expected $q$ values for these disks based on the $\beta$ value from the modified blackbody fits. Rearranging the relation above we obtain, $q = 3 + (\beta/\beta_{s})$. For HD~105, the best-fit value of $\beta = 0.30~\pm~0.05$. From this we calculate that $q = 3.17~\pm~0.03~^{+0.10}_{-0.04}$, assuming a value of $\beta_{s} = 1.8$ and a systematic component to the uncertainty coming from the range of possible values of $\beta_{s}$. Similarly, in the case of HD~131835 we find that $\beta = 0.50~\pm~0.05$ and correspondingly $q = 3.28~\pm~0.03~^{+0.14}_{-0.06}$.}

\begin{table*}
\caption{Grain size distribution slopes, $q$. \label{tab:qvals}}
\centering
\begin{tabular}{lccccccc}
\hline
Target & $\lambda_{\rm (sub-)mm}$ & Instrument & Flux  & Reference & $\alpha_{\rm mm}$ & $\alpha_{\rm Pl}$ & $q$ \\
       &        (mm)              &            & (mJy) &           &                   &                   &     \\
\hline\hline
HD~105    &   0.87 & APEX   & 10.7~$\pm$~5.9 & 1 & 2.41~$\pm$~0.16 & 1.94~$\pm$~0.07 & 3.26~$\pm$~0.17~$^{+0.10}_{-0.03}$ \\  
HD~20807  &   0.50 & SPIRE  & 20.3~$\pm$~7.7 & 2 & 2.05~$\pm$~0.10 & 1.79~$\pm$~0.28 & $>$3.15 \\
HD~61005  &   {1.30} & {SMA} & {\phantom{0}7.2~$\pm$~1.5} & 3 & 2.48~$\pm$~0.08 & {2.05~$\pm$~0.05} & {3.24~$\pm$~0.09~$^{+0.09}_{-0.02}$} \\
HD~109085 &   0.85 & SCUBA2 & 14.3~$\pm$~1.1 & 4 & 2.10~$\pm$~0.07 & 1.90~$\pm$~0.12 & $>$3.11 \\
HD~109573 &   0.85 & SCUBA2 & 14.4~$\pm$~1.9 & 4 & 2.73~$\pm$~0.10 & 1.98~$\pm$~0.02 & $>$3.42 \\
HD~131835 &   0.87 & APEX   & \phantom{0}8.5~$\pm$~4.4 & 5 & 2.17~$\pm$~0.13 & 1.95~$\pm$~0.05 & 3.12~$\pm$~0.14~$^{+0.05}_{-0.01}$ \\
\hline
\end{tabular}

\medskip
\raggedright

{Notes:} Values of $\alpha_{\rm Pl}$ and $\alpha_{\rm mm}$ for targets with non-detections at 9~mm were calculated based on the 3-$\sigma$
 upper limit.

{References:} 1. \cite{2010Nilsson} ; 2. \cite{2013Eiroa} ; 3. \cite{2013Ricarte} ; 4. Holland et al. (in prep.); 5. \cite{2015Moor}.

\end{table*}

\section{Discussion}\label{sec:Dis}

\subsection{Individual results}

\subsubsection{HD~105}

HD~105 is a young G0~V star at a distance of 40~pc that hosts a moderately bright debris disk ($L_{\rm dust}/L_{\star} \sim 2.6\times10^{-4}$). As a member of the Tucana-Horologium association \citep{2000Torres,2000ZucWeb} the system's age of 30~Myr is reasonably well determined. Its debris disk was marginally resolved in far-infrared \textit{Herschel}/PACS imaging observations and found to have a radius of 52~au \citep{2012Donaldson}. HD~105 might therefore be considered an analogue of the young Sun. {The solitary sub-millimetre photometric point at 870~$\mu$m from APEX \citep{2010Nilsson} only provides weak constraint on the disk sub-millimetre emission, and lies 1-$\sigma$ above the best-fit model obtained with our new 9-mm flux measurement of 42~$\pm$~14~$\mu$Jy.} A value of $q = 3.26~\pm~0.17$ was derived for the grain size distribution exponent for this disk. 

\subsubsection{HD~20807}

HD~20807 is another nearby Sun-like star (G1~V), lying at 12~pc. {It was identified as a debris disk host star and thought to be resolved by} \textit{Herschel}/PACS observations \citep{2013Eiroa}. {The putative} disk is thought to be non-axisymmetric, potentially due to the presence of a sub-stellar companion on a wide orbit \citep{2014Faramaz}. \textit{Herschel}/SPIRE measurements in the sub-millimetre reveal a flat spectrum out to $\sim 500~\mu$m, offering no clues as to the point of turnover or steepness of the decline in disk emission. {Subsequent analysis of the same \textit{Herschel} data have cast doubt upon the interpretation of HD~20807 as a debris disk host star, with \cite{2015MM} finding no evidence for a far-infrared excess.} We obtained a 3-$\sigma$ upper limit of 54~$\mu$Jy/beam from the ATCA measurements, and a corresponding lower limit of $q > 3.15$. The beam size of ATCA in compact mode and \textit{Herschel}/PACS are comparable, so the non-detection of this target may be due to dilution of the total disk flux over multiple beams {or, potentially, due to the absence of any disk around the star in the first place}. 

\subsubsection{HD~61005}

HD~61005 (the moth) is a very well studied young debris disk system around a late G-type star at a distance of 35~pc. Its disk is bright, having been imaged in scattered light \citep{2007Hines}, and resolved at millimetre wavelengths \citep{2013Ricarte,2016Macgregor}. The disk exhibits a pronounced asymmetry that has been attributed to either interaction with the interstellar medium \citep{2009Maness}, or the presence of planetary companion perturbing the disk \citep{2010Buenzli,2016Esposito}. We obtained a 9-mm flux density of 59~$\pm$~14~$\mu$Jy for this system. This is completely consistent with the 9-mm flux obtained by \cite{2016Macgregor}. Using our 9-mm measurement in combination with the SCUBA-2 flux density at 850~$\mu$m we obtained $q = 3.08~\pm~0.17$ for the disk. This is marginally steeper than the earlier derivation of \cite{2016Macgregor} ($q~=~3.32~\pm~0.06$) using the 1.3-mm SMA flux density (7.2~$\pm$~1.5~mJy) to calculate $\alpha_{\rm mm}$ {, for which we obtain $q~=~3.24~\pm~0.09$ (as presented in Table \ref{tab:qvals}.}

\subsubsection{HD~109085}

HD~109085 ($\eta$ Crv), a 1.5~Gyr-old F2~V star at a distance of 55 pc, hosts a disk with {an inner warm component that is anomalously bright for its age}. The disk was resolved into two distinct components by \textit{Herschel}/PACS \citep{2014Duchene}. The outer belt was recently resolved at millimetre wavelengths by ALMA observations that also identified CO emission from the inner regions of the disk \citep{2017Marino}. This places it (along with HD~131835) amongst the small number of debris disks that also have detectable gas \citep{2016Greaves}. Interpretation of its mid-infrared spectrum revels evidence of primordial material, believed to originate from icy planetesimals scattered from the outer belt to the inner regions of the system \citep{2012Lisse}. We obtained an upper limit of 36~$\mu$Jy/beam with ATCA. Due to the extended nature of the source, covering an area of $\sim$ 3 beams, the {outer belt's} surface brightness is too low to be detectable with the sensitivity of ATCA. However, our measurement was a meaningful constraint to the outer disk as resolved in \cite{2017Marino}. We thus obtained a lower limit of $q > 3.11$; this value was calculated taking into account the area of the disk. 

\subsubsection{HD~109573}

{HD~109573 (HR~4796 A), an 8~Myr old A-type star at 73~pc, is the most distant of our targets. It was the second disk to be spatially resolved (after $\beta$ Pictoris) being imaged at mid-infrared wavelengths \citep{1998Jayawardhana,1998Koerner}. The disk has also been imaged successfully in scattered light, being dubbed the `eye of Sauron' \citep{2009Schneider}. The disk has a narrow radial width \citep[$\Delta$R/R $\sim$ 14 per cent][]{2015Rodigas} and its structure has been interpreted as indicating the presence of a shepherding planetary companion \citep{2011Thalmann}.} Given the disk's emission is well-sampled, and its small angular extent, we expected that this system would yield a clear detection in our 9-mm observations. However, we only obtained a 3-$\sigma$ upper limit of 63~$\mu$Jy/beam equating to a limit on $q > 3.42$. This is well within the expected range of 3 to 4 for debris disks. {We attribute the non-detection of this system to the elevated millimetre photometry from APEX \citep[21.5~$\pm$~6.6~mJy][]{2010Nilsson} and JCMT/SCUBA \citep[19.1~$\pm$~3.4~mJy,][]{2000Greaves} compared to the more recent JCMT/SCUBA2 measurement (14.4~$\pm$~1.9~mJy, Holland in prep.) that biased the modified blackbody fit.}

\subsubsection{HD~131835}

HD~131835 is a young (17~Myr old), massive analogue to the $\beta$ Pictoris system. Sub-millimetre observations of the disk detected the presence of CO gas in the disk \citep{2015Moor}, making it only one of a handful of such systems \citep[see e.g.][]{2016Greaves}. It has recently been imaged in scattered light by GPI, revealing a broad (75 to 210 au), inclined (75$\degr$) disk around the host star \citep{2015Hung}. The disk was also marginally resolved by \textit{Herschel}/PACS far-infrared observations, suggesting an extent of 170~au \citep{2015Moor}. Here we obtained the first measurement of the disk continuum emission 9 mm, with a flux of 53~$\pm$~17~$\mu$Jy. The corresponding $q = 3.12~\pm~0.14$. This disk has the shallowest sub-millimetre slope amongst the disks measured here; this could be attributed to the influence of gas in the disk moderating the collisions between dust grains.

\subsection{ {Comparision with previous work} }

{The $q$ values obtained for the newly detected disks were smaller than the average value for mm-detected disks calculated in \cite{2016Macgregor} (3.36~$\pm$~0.02), and that of the commonly-adopted steady state collisional cascade of \cite{1969Dohnanyi}. Combining values for the three detected disks in the sample we found an error-weighted $q$ of 3.15~$\pm$~0.09. If we add the three non-detections (treating the 3-$\sigma$ upper limit at 9-mm as a detection), the mean $q$ increases to 3.26~$\pm$~0.05, and is still smaller than the value of \cite{2016Macgregor}. The lower sensitivity of ATCA compared to VLA at the same wavelength would account for our detection of disks with smaller $q$ values, as these disks have a greater contribution from larger grains to their total emission and would therefore be relatively brighter.}

{At first glance the values are also discrepant with those obtained by \cite{2012Gaspar}. An assumption of that work was that millimetre emission from the disk is dominated by grains with $Q_{\rm abs} = 1$, such that the relationship between $q$ and $\alpha_{\rm mm}$ is $q = \alpha_{rm mm} + 1$. This is contrary to the results of \cite{2006Draine}. We have recalculated the values of $q$ for the nine disks examined in \cite{2012Gaspar} using the same method employed here and in \cite{2016Macgregor}. A side-by-side comparison of the derived values for $q$ using the same photometry are given in Table \ref{tab:q_comp}.}

\begin{table}
\caption{Comparison of $q$ values derived by the method of \cite{2012Gaspar} and in this work. \label{tab:q_comp}}
\centering
\begin{tabular}{lcc}
\hline
Target         & $q_{\rm Gaspar}$ & $q_{\rm This work}$ \\
               &                  &                     \\
\hline\hline
$\beta$ Pic    &  3.63$^{+0.24}_{-0.26}$  &  3.53~$\pm$~0.10  \\
$\epsilon$ Eri &  3.14$^{+0.49}_{-0.14}$  &  2.53~$\pm$~0.18  \\
$\alpha$ PsA   &  3.62$^{+0.26}_{-0.20}$  &  3.21~$\pm$~0.07  \\
HD~8907        &  3.00$^{+0.56}_{-0.00}$  &  2.64~$\pm$~0.20  \\
HD~104860      &  3.07$^{+0.57}_{-0.07}$  &  2.70~$\pm$~0.22  \\
HD~107146      &  3.34$^{+0.42}_{-0.34}$  &  3.11~$\pm$~0.12  \\
HR~8799        &  3.21$^{+0.33}_{-0.21}$  &  4.11~$\pm$~0.33  \\
$\alpha$ Lyr   &  4.01$^{+0.30}_{-0.28}$  &  3.61~$\pm$~0.17  \\
HD~207129      &  3.42$^{+0.65}_{-0.42}$  &  3.74~$\pm$~0.38  \\
\hline
\end{tabular}

\medskip
\raggedright

\end{table}

{In all cases except two, the values of $q$ are found to be smaller using the method applied here and in \cite{2016Macgregor} than by that of \cite{2012Gaspar}. In combination, the revised disk measurements have a mean $q = 3.23~\pm~0.04$. The revised values we obtain are more in line with the findings of \cite{2016Macgregor} and numerical results \citep[e.g.][]{2014Schuppler,2015Schuppler} than with the modelling results of far-infrared excesses \citep{2014Pawellek}. In the two instances where our method yields a steeper $q$ than Gaspar et al., i.e. HR~8799 and HD~207129, we note the longer wavelength of the sub-millimetre SED used to calculate $\alpha_{\rm mm}$ is poorly constrained, being represented by sub 2-$\sigma$ detections in each case; better quality sub-millimetre photometry would likely yield a flatter size distribution (smaller $q$).}

{Three of the disks, $\epsilon$ Eri, HD~8907, and HD~104860, have $q$ values $< 3$. These are outside the generally expected range of 3 to 4 for debris disks, and beyond the regime where the relations derived in \cite{2006Draine} were found to hold. In \cite{2016Macgregor} they identified one system, HD~141569, with similar properties ($q = 2.84~\pm~0.05$). They attributed its small size distribution exponent to the fact that it was a young, gas-rich protoplanetary disk system unlike the debris disks that made up the majority of their sample. Here, such a line of reasoning does not hold as all three are well-known debris disk systems.}

{We therefore investigated the impact of more recent, higher quality photometry than that presented in \cite{2012Gaspar}. Using recent millimetre photometry from the SMA \citep{2016Steele} and LMT \citep{2016Chavez}, we recalculated the values of these three disks and found them consistent with the remainder of the sample, with $q = 3.21~\pm~0.31$ ($\epsilon$ Eri), $3.69~\pm~0.23$ (HD~8907), and $3.26~\pm~0.59$ (HD~104860). With the revised measurements, the mean $q$ value of Gaspar et al.'s sample changes to $3.35~\pm~0.05$, completely consistent with \cite{2016Macgregor}.}

\section{Conclusions}\label{sec:Con}

We set out to measure the continuum emission from six debris disks at millimetre wavelengths using the ATCA in order to constrain the slope of their submillimetre emission and thereby better determine the planetesimal properties in those systems. Our observations have demonstrated the capability of ATCA to detect continuum emission from relatively faint debris disks ($L_{\rm dust}/L_{\star} < 10^{-3}$) that are most representative of the bulk of known, cool circumstellar disks. 

We detected three of the six disks at s/n $>$ 3 in our observations. Two of the targets, HD~105 and HD~131835, were detected for the first time at millimetre wavelengths. HD~61005 had been previously observed at 9 mm with a comparable flux density to the measurement presented here. Upper limits for the remaining three targets (HD~20807, HD~109085, and HD~109573) placed strong constraints on the slope of the disks' sub-millimetre emission. 

We found that the exponents of the size distribution, $q$, for the three detected disks were smaller than both the steady state collisional cascade model of \cite{1969Dohnanyi}, and also smaller than the mean $q$ value obtained for the sample of 15 debris disks examined in \cite{2016Macgregor}. For the detected disks, the $q$ values are most consistent with theoretical models wherein the planetesimals are treated as `rubble piles' without any rigid strength \citep{2005PanSar}, and for the most part consistent with collisional modelling results of \cite{2012Lohne} and \cite{2014Schuppler,2015Schuppler}. The {lower} limits on the $q$ values of HD~20807 and HD~109085 are both high enough that we cannot rule out consistency with \cite{2016Macgregor}, whilst the upper limit on HD~109573's $q$ value is low enough such that the disk grain size distribution must be steeper than is typical for millimetre-detected disks. Given the brightness of HD~109573's disk at sub-millimetre wavelengths and its relatively compact nature, further study at comparable wavelengths and higher sensitivity (e.g. 9-mm observations on the Jansky VLA) should be expected to yield a detection of the disk.

\section*{Acknowledgments}

{The authors would like to thank the referee for their careful and considerate comments that improved the manuscript}. JPM would like to thank Paul Jones for helpful discussions on the ATCA data reduction process. JPM is supported by a UNSW Vice-Chancellor's Postdoctoral Research Fellowship. This research has made use of the SIMBAD database, operated at CDS, Strasbourg, France. This research has made use of NASA's Astrophysics Data System. The Australia Telescope Compact Array is part of the Australia Telescope National Facility which is funded by the Australian Government for operation as a National Facility managed by CSIRO.

\noindent \textit{Facilities:} ATCA

\bibliographystyle{mnras}
\bibliography{platypus}

\bsp	

\label{lastpage}
\end{document}